\documentclass[preprint,12pt]{elsarticle}
\usepackage{graphicx}
\usepackage{amssymb}
\usepackage{amsthm}
\usepackage{amsmath}
\usepackage{xcolor}
\usepackage{booktabs}
\usepackage{multirow}

\journal{}

\begin{document}

\begin{frontmatter}


\title{In-situ On-demand Digital Image Correlation: A New Data-rich Characterization Paradigm for Deformation and Damage Development in Solids}



\author{Ravi Venkata Surya Sai Mogilisetti}
\author{Partha Pratim Das}
\author{Rassel Raihan}
\author{Shiyao Lin \corref{cor1}}

\address[add1]{University of Texas at Arlington, Arlington, TX, 76010}
\address[add2]{Institute for Predictive Performance Methodologies, Fort Worth, TX, 76118, USA}

\cortext[cor1]{Corresponding author  shiyao.lin@uta.edu (Shiyao Lin)}

\begin{abstract}
Digital image correlation (DIC) has become one of the most popular methods for deformation characterization in experimental mechanics. DIC is based on optical images taken during experimentation and post-test image processing. Its advantages include the capability to capture full-field deformation in a non-contact manner, the robustness in characterizing excessive deformation induced by events such as yielding and cracking, and the versatility to integrate optical cameras with a variety of open-source and commercial codes. In this paper, we developed a new paradigm of DIC analysis by integrating camera control into the DIC process flow. The essential idea is to dynamically increase the camera imaging frame rate with excessive deformation or deformation rate, while maintaining a relatively low imaging frame rate with small and slow deformation. We refer to this new DIC paradigm as in-situ on-demand (ISOD) DIC. ISOD DIC enables real-time deformation analysis, visualization, and closed-loop camera control. ISOD DIC has captured approximately 178\% more images than conventional DIC for samples undergoing crack growth due to its dynamically adjusted frame rate, with the potential to significantly enhance data richness for damage inspection without consuming excessive storage space and analysis time, thereby benefiting the characterization of intrinsic constitutive behaviors and damage mechanisms. 

\end{abstract}


\end{frontmatter}


\section{Introduction}
\label{Section1}

Digital image correlation (DIC) is a versatile, robust, and effective deformation characterization approach for experimental solid mechanics. Ever since the development of its core algorithm and the commercialization of the technique (e.g., GOM Correlate, VIC-3D, MatchID, etc.), DIC has been applied to a wide range of industry sectors, including aerospace \cite{lin2021effect,lin2022experimental,lin2020predicting}, civil \cite{fan2025civil,liu2025civil}, sporting \cite{ghaednia2017sport}, bioengineering \cite{seamone2025eyelid, sugerman2023speckling}, geography \cite{caporossi2018digital}, and art conservation \cite{malowany2014application}. The unique advantage of DIC is that, compared to more conventional strain-gauge-based measurement \cite{huang2024straingauge,ajovalasit2011strainGauge} and shadow Moiré measurement \cite{ding2002moire,rakow2005moire}, DIC is non-contact, significantly easier to operate, and able to provide full-field deformation measurement. Implemented with high-speed cameras, high-speed DIC is capable of acquiring highly detailed insights for high-speed impact events, which are hardly characterizable by conventional methods.

Most DIC programs can be divided into two categories: local subset-based DIC and global finite-element (FE)-based DIC \cite{localGlobalDIC1, localGlobalDIC2,localGlobalDIC3}. In local subset-based DIC, each image is mapped to multiple subsets, and the deformation of each subset is computed independently. Because computations are parallelized across subsets, subset-based DIC can be highly efficient \cite{WANG2016200}. However, since the solutions are performed on subsets separately, the deformation across subset boundaries may be incompatible. In global FE-based DIC, the global deformation field is tracked as a whole set, resulting in lower noise and improved deformation compatibility, but reduced efficiency. In recent years, optical flow, a robust motion analysis model prevalently used in computer vision, has emerged as a new deformation measuring approach for solids \cite{deng2020novelOpticalFlow, bauer2023spatioOpticalFlow, zhang20242dOpticalFlow}. The accuracy and efficiency of conventional DIC and optical flow have been systematically compared in \cite{OpticalFlowVsDIC1, OpticalFlowVsDIC2}. The Lucas–Kanade (LK)-based optical flow model \cite{LKOriginal} has been found capable of providing similar results and accuracy levels of deformation measurement compared to DIC. Optical flow is generally faster and more suitable for real-time characterization applications. However, one noticeable disadvantage is that optical flow models are intrinsically sensitive to the variation in lighting conditions, posing challenges for measurements done in optically harsh environments (e.g., welding, high temperature, etc.). 

Besides commercial DIC packages, open-source DIC codes have also been developed and applied. Ncorr \cite{blaber2015ncorr} was developed as a Matlab-based open-source subset-based 2D DIC package that amalgamates modern DIC algorithms and has been adopted by many \cite{KUMAR2019100061}. OpenCorr was developed as an open-source C++ library, providing full-function modules of 2D and stereo DIC, as well as digital volume correlation (DVC) \cite{jiang2023opencorr}. RealPi2dDIC \cite{das2021realpi2ddic} enables a low-cost solution by integrating Raspberry Pi as the computation core and the Pi camera as the imaging device. Furthermore, RealPi2dDIC is capable of real-time full-field deformation measurement. A series of works done by Yang et al. \cite{yang2019ALDIC,yang2020ALDVC,yang2021fast} develops a new paradigm for DIC and DVC by combining the advantages of both the local subset-based DIC and global finite-element-based DIC to achieve an improved balance between accuracy and efficiency. Recently, deep learning models have been integrated into the DIC workflow, enabling accurate and ultra-fast deformation measurements \cite{yang2022deep,boukhtache2021deep}. The deep learning models were trained on synthesized datasets and replaced the subset-based DIC computation solver, resulting in significantly improved analysis efficiency.

In addition to serving as virtual strain gauges, DIC has been recently integrated into the process of material characterization. A novel approach was proposed to enable the characterization of the constitutive relationship of materials directly from the raw data consisting of images with DIC and force measurement \cite{wihardja2025constitutive}. Two-dimensional heterogeneous materials were characterized with full-field strain data enabled by DIC in conjunction with the finite element method (FEM) \cite{kattil2025heterogeneous}. In addition to single-modality characterization, DIC has also been applied together with other sensors, such as thermal cameras \cite{cholewa2016technique, tuo2022study} and acoustic emission sensors \cite{DIC_AE1, DIC_AE2, DIC_AE3}. Such multi-modal characterizations have enabled deepened insights into the fundamental deformation and damage behaviors of novel materials. 

In this paper, we propose a new paradigm of DIC-based deformation measurement with DIC cameras in the loop. The new paradigm provides in-situ on-demand (ISOD) DIC measurement capabilities, enabling data-rich characterization for the deformation and damage of tested samples. Essentially, ISOD DIC cohesively integrates camera control into the in-situ characterization loop, allowing for increased camera frame rate when the deformation becomes large or fast. When a local high deformation gradient or accelerated rate of deformation is captured, complex responses such as material yielding and fracturing are likely to happen. Therefore, ISOD DIC increases the characterization information richness by increasing the frame rate of imaging correspondingly. These localized events are usually highly valuable for the purposes of characterizing intrinsically complicated behaviors such as strain hardening and damage initiation/propagation. 

In Section 1, a brief introduction of DIC and the key concept of ISOD DIC are provided. In Section 2, the core analysis modules of ISOD DIC, optical flow, and the ISOD DIC process flow are outlined. Results and discussions are provided in Sections 3 and 4, starting with a validation example on the uniaxial testing of aluminum dog-bone samples. Then, the ISOD DIC characterization results for additively manufactured biomimetic samples are presented, demonstrating ISOD DIC's effectiveness in enriching characterization information during complex damage events. The conclusions are outlined in Section 5.
\section{Methodology}
\label{Section2}
In this paper, we developed the 2D ISOD DIC by adopting an optical flow model based on the Lucas–Kanade (LK) algorithm \cite{LKOriginal} for its agility to achieve real-time characterization. In this section, the theoretical background of the optical flow model is introduced first, followed by a high-level illustration of the process flow of the ISOD DIC. 

\subsection{ISOD DIC Analysis Core Module: Optical Flow}
Optical flow is a technique that analyzes the motion of pixels in a sequence of images. The algorithm helps in calculating a 2D vector field of deformation that describes the motion of each pixel. Based on the field of deformation, strain fields are calculated. 

\subsubsection{LK-based Optical Flow Algorithm}
Consider two consecutive images for a deformed solid with a time difference of $\Delta t$. The two images can be described in terms of the light intensity fields $I(x, y, t)$, where $(x,y)$ refer to pixel coordinates. LK-based optical flow is performed on grayscale intensity values rather than color. Therefore, when the color images are captured, they need to be converted into grayscale before computing the optical flow. To increase the computational speed, all the images can be captured in grayscale format.   With the deformation, light intensity should be maintained as in Equation \ref{Eqn1}.

\begin{equation}
I(x, y, t) = I(x + u, y + v, t + \Delta t)
\label{Eqn1}
\end{equation}

Applying Taylor's Series expansion on the right-hand side,

\begin{equation}
I(x + u, y + v, t + \Delta t) = I(x, y, t) + I_x u + I_y v + I_t \Delta t
\label{Eqn2}
\end{equation}

\noindent where,
\[
I_x = \frac{\partial I}{\partial x}, \quad 
I_y = \frac{\partial I}{\partial y}, \quad 
I_t = \frac{\partial I}{\partial t}
\] 
represent the image intensity gradients with respect to \(x\), \(y\), and \(t\), respectively. Here, \(u\) and \(v\) denote the displacements in the \(x\) and \(y\) directions.

From Equations \ref{Eqn1} and \ref{Eqn2},

\begin{equation}
    I(x, y, t) = I(x, y, t) + I_x u + I_y v + I_t \Delta t
    \label{eqn3}
\end{equation}

\begin{equation}
    I_x u + I_y v + I_t = 0
    \label{eqn4}
\end{equation}

Equation \ref{eqn4} is known as the optical flow constraint equation.

\begin{figure}[htbp]
    \centering
    \includegraphics[width=0.8\textwidth]{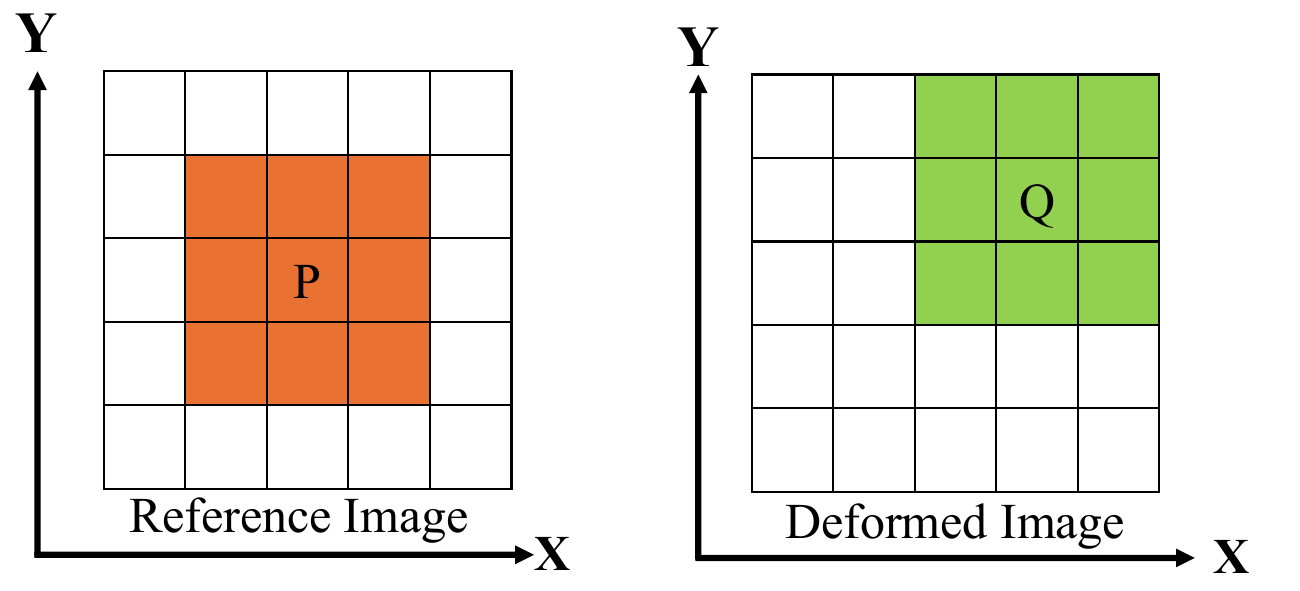}
    \caption{Schematic representation for displacement calculation of a pixel.}
    \label{fig:REF_VS_DEF}
\end{figure}

To evaluate a displacement for each pixel, surrounding pixels must also be considered. This is because \ref{eqn4} consists of 2 unknown variables, which makes a problem in estimating the \(x\), \(y\). The LK-based optical flow algorithm assumes the same intensity of neighboring pixels. A 3$\times$3 patch surrounding each pixel, assuming that the motion $(u, v)$ is constant throughout the patch, as shown in Figure \ref{fig:REF_VS_DEF}. However, the 3$\times$3 patch size is not fixed and can change depending on factors such as noise level, resolution, and predicted deformation. The displacement is calculated using numerical differentiation, as represented in Equations \ref{Eqn5} - \ref{Eqn8}. In the equations, the indices 1 to 9 correspond to the nine pixels of the 3$\times$3 patch. $\mathbf{A}$ and $\mathbf{B}$ are spatial and temporal gradient matrices of the light intensity.

\begin{equation}
\begin{cases}
I_{x1}u + I_{y1}v = I_{t1} \\
I_{x2}u + I_{y2}v = I_{t2} \\
\vdots \\
I_{x9}u + I_{y9}v = I_{t9}
\end{cases}
\qquad
\label{Eqn5}
\end{equation}

\begin{equation}
\begin{bmatrix}
I_{x1} & I_{y1} \\
I_{x2} & I_{y2} \\
\vdots & \vdots \\
I_{x9} & I_{y9}
\end{bmatrix}
\begin{bmatrix}
u \\ v
\end{bmatrix}
= -
\begin{bmatrix}
I_{t1} \\ I_{t2} \\ \vdots \\ I_{t9}
\end{bmatrix}
\qquad 
\label{Eqn6}
\end{equation}

\begin{equation}
\mathbf{A}
\begin{bmatrix}
u \\ v
\end{bmatrix}
= \mathbf{B} 
\qquad
\label{Eqn7}
\end{equation}

\begin{equation*}
\mathbf{A}^{T} \mathbf{A}
\begin{bmatrix}
u \\ v
\end{bmatrix}
= \mathbf{A}^{T} \mathbf{B}
\end{equation*}

\begin{equation}
\begin{bmatrix}
u \\ v
\end{bmatrix}
= (\mathbf{A}^{T} \mathbf{A})^{-1} \mathbf{A}^{T} \mathbf{B}
\qquad
\label{Eqn8}
\end{equation}

The displacement vector calculated using the LK algorithm is used to calculate Green strains for every pixel in the region of interest (ROI) following Equations \ref{Eqn9}-\ref{Eqn11}.

\begin{equation}
\varepsilon_{xx} = 
\frac{\partial u}{\partial x} 
+ \frac{1}{2} 
\left[
\left( \frac{\partial u}{\partial x} \right)^2 
+ \left( \frac{\partial v}{\partial x} \right)^2
\right]
\label{Eqn9}
\end{equation}

\begin{equation}
\varepsilon_{yy} = 
\frac{\partial v}{\partial y} 
+ \frac{1}{2} 
\left[
\left( \frac{\partial u}{\partial y} \right)^2 
+ \left( \frac{\partial v}{\partial y} \right)^2
\right]
\label{Eqn10}
\end{equation}

\begin{equation}
\varepsilon_{xy} = 
\frac{1}{2}
\left(
\frac{\partial u}{\partial y}
+ \frac{\partial v}{\partial x}
+ \frac{\partial u}{\partial x} \frac{\partial u}{\partial y}
+ \frac{\partial v}{\partial x} \frac{\partial v}{\partial y}
\right)
\label{Eqn11}
\end{equation}

\subsubsection{Optical Flow vs. Conventional DIC}
Although the accuracy and efficiency have been systematically compared in \cite{OpticalFlowVsDIC1, OpticalFlowVsDIC2}, a brief conceptual comparison between optical flow and DIC is still drawn here. Conventional subset-based DIC mainly works on cross-correlation or least-squares matching of the subsets between the reference and the deformed image. The local displacement vectors $(u,v)$ are determined by finding the position of the subset in the deformation image that closely matches the reference image. Another commonly used DIC method is finite element (FE)-based DIC, which is also referred to as global DIC. FE-based DIC estimates the deformation field over the ROI by using the finite element shape function. Instead of correlating each subset of pixels, the strain field is represented using nodal degrees of freedom. This approach helps in reducing the optimization issues between the reference image and deformation image, which helps in providing smooth displacement and strain histories.  

In this work, LK-based optical flow is used as a theoretical base and computational core to perform DIC. The LK-based optical flow algorithm, as outlined in Section 2.1.1, is derived based on the constant brightness, which means that the light intensity remains constant during the test from the reference image to the last image. Optical-flow-based DIC enables dense, full-field displacement estimation and higher spatial resolution because it tracks each pixel. To generate accurate full-field displacement across the ROI, optical-flow-based DIC considers the deformation of the pixels surrounding it, as mentioned in Figure \ref{fig:REF_VS_DEF}.

One prominent disadvantage of conventional DIC models is the relatively larger time consumption compared to optical-flow-based DIC, preventing their applications in in-situ characterization. Optical flow, with significantly faster computation efficiency and comparable accuracy \cite{OpticalFlowVsDIC1, OpticalFlowVsDIC2}, has been used as a new DIC computation core \cite{bauer2023spatioOpticalFlow, deng2020novelOpticalFlow, zhang20242dOpticalFlow}. However, optical flow is sensitive to environmental light intensity change, requiring a constant illuminating condition for the testing environment. In this paper, all the experimental tests were done with constant lighting. 

\subsection{ISOD DIC Process Flow}
\begin{figure}[htbp]
    \centering
    \includegraphics[width=1\textwidth]{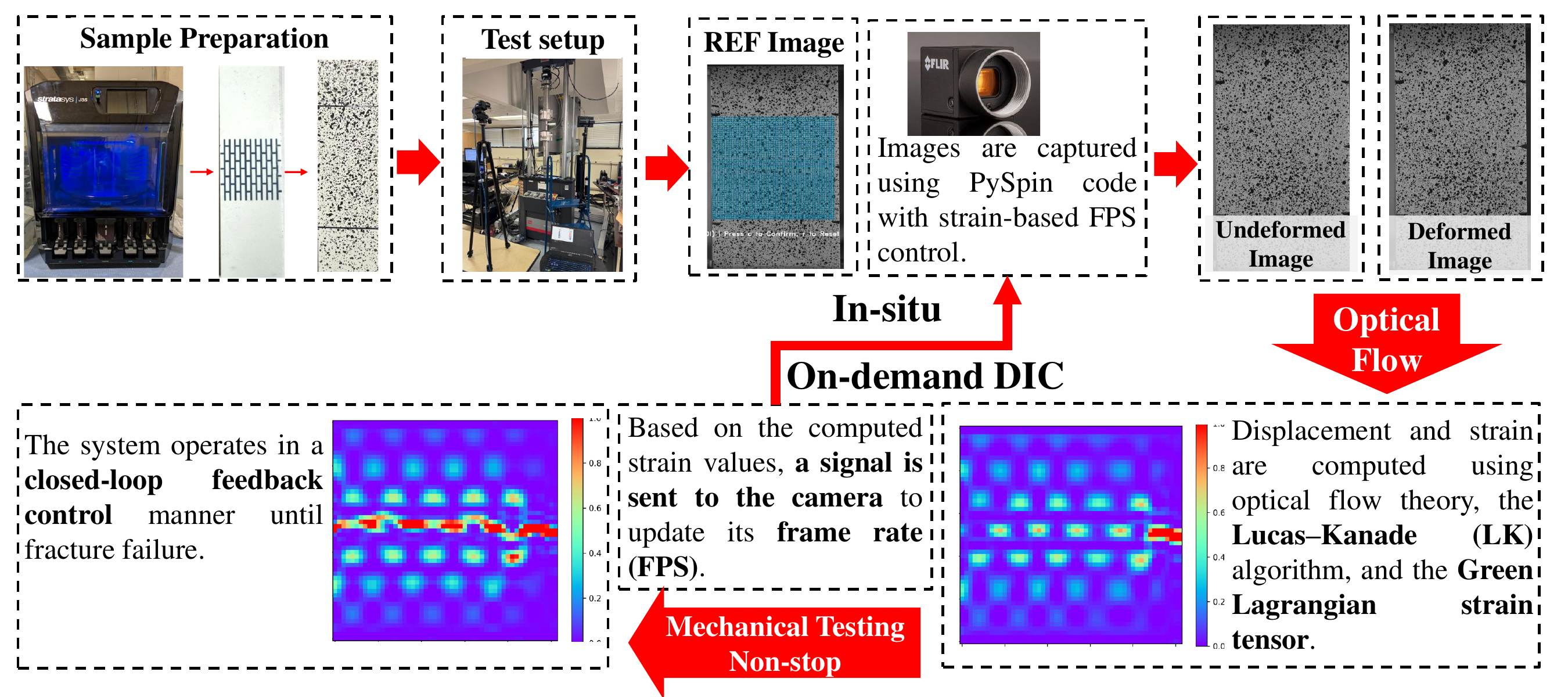}
    \caption{Workflow of ISOD DIC.}   
    \label{fig_ISOD_workflow}
\end{figure}

The experiment setup and ISOD DIC process flow are outlined in Figure \ref{fig_ISOD_workflow}. It should be noted again that the ISOD DIC characterizations in this paper are all 2D deformation characterizations using one machine vision camera. As outlined in Figure \ref{fig_ISOD_workflow}, an ROI is defined in the first captured image. As the mechanical test continues, LK-based optical flow is applied to the image sequence, from which the displacement and strain fields are calculated. Based on the computed strain values, a signal is sent to the machine vision camera to dynamically adjust its image-capturing frame rate until the completion of the test. More details of image capturing, deformation, strain field computation, camera frame rate adjusting, and hardware and software integration can be found in Sections 2.2.1 - 2.2.4.

\label{sec2.2} 

\subsubsection{Batch-wise Image Capturing}
\begin{figure}[htbp]
    \centering
    \includegraphics[width=1\textwidth]{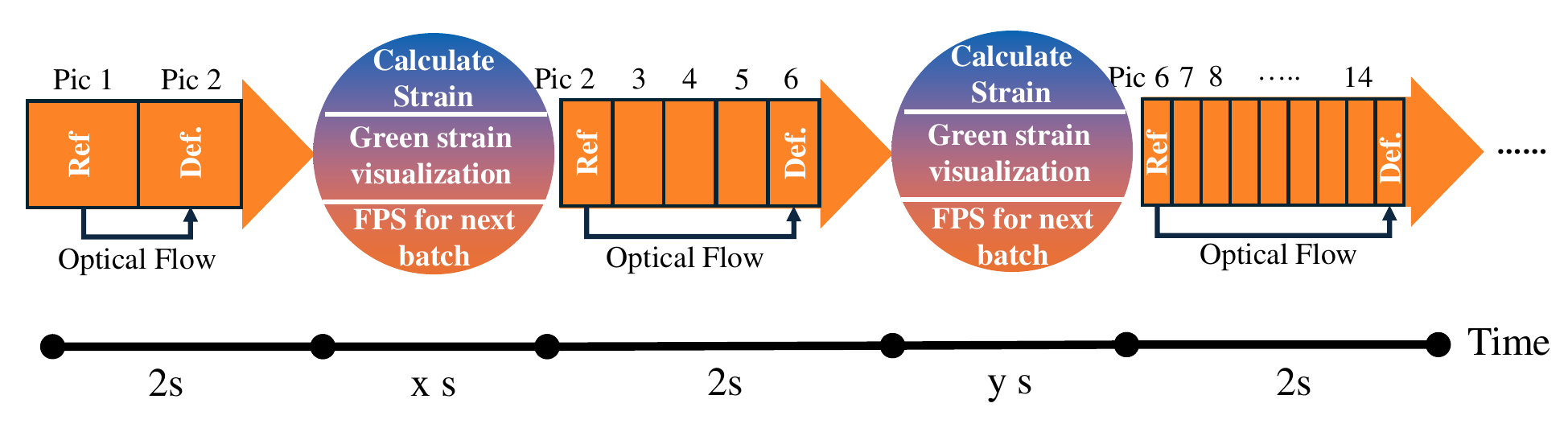}
    \caption{Batch-wise feedback-controlled ISOD DIC characterization.}   
        \label{fig_Batch_wise_DIC}
\end{figure}

In the ISOD-DIC process, image capturing and strain field computation are performed in a batch-wise, feedback-controlled manner to optimize data capture during the initial Stage of the test and computational efficiency. Figure \ref{fig_Batch_wise_DIC} describes the batch-wise data acquisition.

After the capturing of the first image, the ROI where the DIC needs to be performed is selected, and the coordinates of the pixels are written in a \textit{.txt} file. Images are captured sequentially for the fixed duration; in this paper, each batch spans 2 seconds. Within each batch, the camera captures a series of images with a constant frame rate (frames per second (FPS)). The frame rate is dynamically adjusted based on the evolving strain in the sample from batch to batch. Therefore, the number of images per batch varies depending on the instantaneous strain value, as illustrated in Figure \ref{fig_Batch_wise_DIC}. As shown in Figure \ref{fig_Batch_wise_DIC}, between each batch, strain calculation, visualization, and camera frame rate adjustment are implemented. As mentioned, each batch of image capturing lasts 2 seconds, whereas the time durations of program running between batches may vary, as $x$ and $y$ seconds in Figure \ref{fig_Batch_wise_DIC}. All images captured during the experiment are systematically stored in the device, which can be analyzed by any commercially available or open source DIC software.

\subsubsection{Deformation Field Computation and Visualization}
At the end of each batch, DIC is performed using an optical flow-based DIC algorithm. Instead of performing DIC for all the images that are captured, DIC is performed between the first and last images of every batch to reduce the computational workload of the program, enabling real-time deformation characterization and visualization. The computed strain values of each batch are used to determine the camera frame rate for the next batch of image acquisition, as shown in Figure \ref{fig_Batch_wise_DIC}. The computed strain fields are stored in \textit{.npz} file, which is a type of zip file for Python Numpy data. The \textit{.npz} file is used to stream both the real-time strain fields when the material is loaded.

\subsubsection{Dynamic Image Capturing Frame Rate Adaptation}
The crucial feature of the ISOD-DIC framework is that it is a closed-loop feedback mechanism for controlling the frame rate of the camera. Based on the strain growth, the camera frame rate is dynamically adjusted.

\noindent\textbf{{Slow and small deformation} $\rightarrow$ {Image acquisition with low frame rate}}

When the value of max strain or strain rate is low, i.e, when the deformation is relatively small, the system captures at a low frame rate to avoid unnecessary data storage and workload.

\noindent\textbf{{Fast and large deformation} $\rightarrow$ {Image acquisition with increased frame rate}}

When the strain value exceeds a pre-determined max strain or strain rate value, the frame rate of the camera is increased to capture more images for the next batch. This ensures more images are captured during the crack initiation. The critical threshold values of strain rates and maximum strains may be determined according to preliminary tests, history test data, and computer simulations. 

\subsubsection{Hardware and Software Integration}
In this paper, a FLIR BlackFly USB 3 camera was used for the ISOD DIC characterization. The FLIR BlackFly USB 3 camera is a machine-vision camera capable of running at a maximum frame rate of 133 Hz and featuring a resolution of 1920 $\times$ 1200. PySpin, an open-source Python wrapper for the Spinnaker software development kit (SDK) for FLIR cameras, was used for controlling the FLIR BlackFly camera. OpenCV \cite{opencv_library}, which is an open-source computer vision library, was utilized to perform the LK-based optical flow algorithm. In addition, Python packages, including Numpy \cite{harris2020array}, SciPy \cite{2020SciPy-NMeth}, and Matplotlib \cite{Hunter:2007} libraries, were utilized to perform numerical analysis, interpolation, and generate the real-time strain field visualization.
 
\begin{figure}[htbp]
    \centering
    \includegraphics[width=0.8\textwidth]{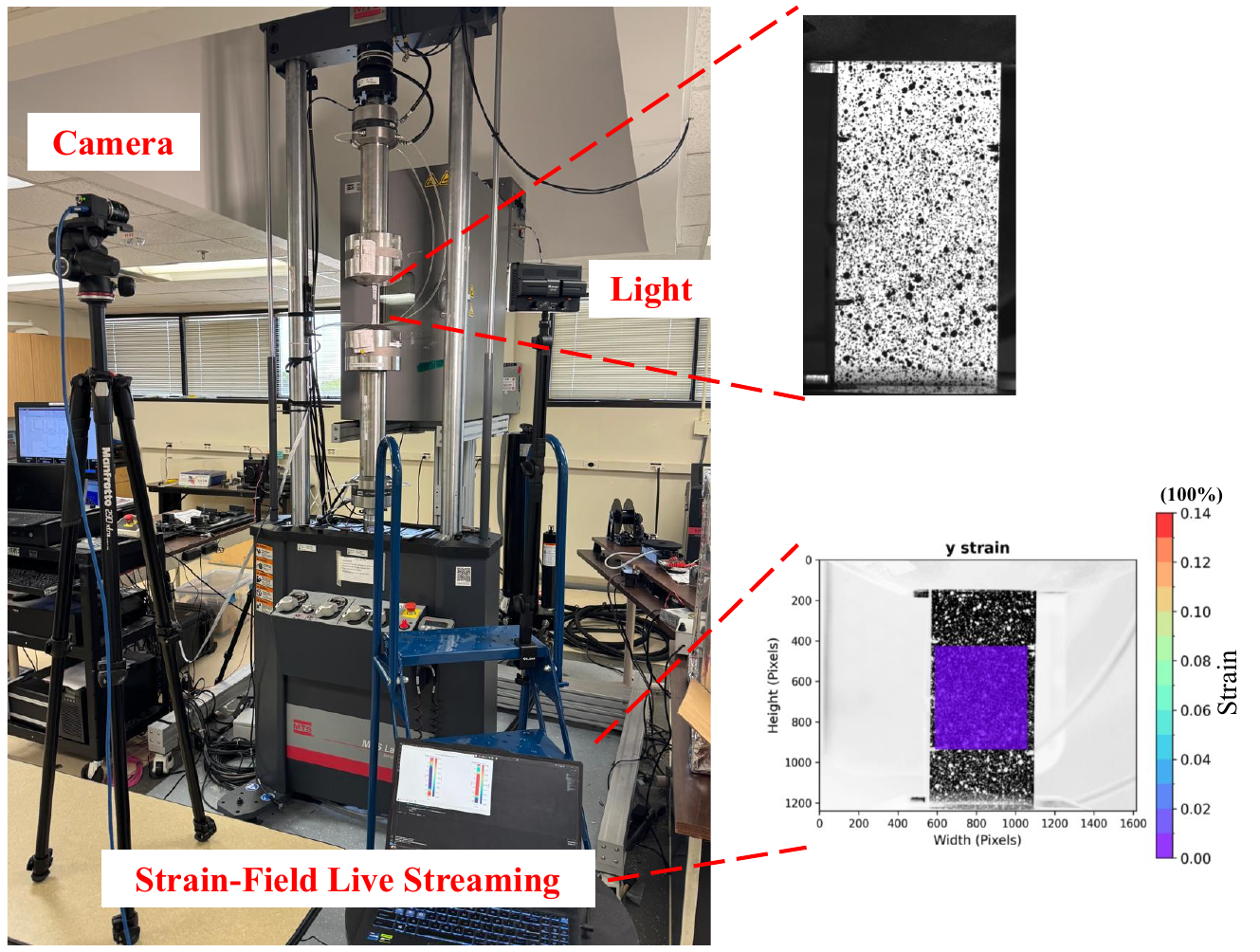}
     \caption{The experimental setup with the ISOD DIC.}
     \label{fig_testSetup}
\end{figure}

 ISOD DIC was performed when the material was under quasi-static uniaxial tensile loading as illustrated in Figure \ref{fig_testSetup}. The camera was connected to a laptop computer with a USB 3 cable. ISOD DIC ran on the laptop, controlling the USB 3 camera to acquire images, compute deformation fields, adjust the camera frame rate, and generate real-time strain fields visualization. An external light source was used to maintain uniform light intensity during the test. For this test, an MSI CYBORG 15 A13V laptop equipped with an Intel i7core processor and Nvidia GeForce RTX 450 GPU  is utilized for high-speed imaging and to perform  ISOD DIC.

\section{Results}
\label{Section3}
In this section, the experimental characterization results are presented to validate the deformation measurement accuracy of ISOD DIC and its effectiveness in enhancing the information richness for damage inspection. 

\subsection{Experimental Cases}
\label{sec3.1}

The ISOD DIC experimental characterization was compared with a commercially available software package, GOM Correlate. There are two experimental cases studied. The first case is the uniaxial tensile test for aluminum dog-bone specimens. This test is a commonly conducted material characterization test for characterizing constitutive laws. This case is selected as the material is homogeneous and isotropic, resulting in smooth and continuous deformation under loading conditions, suitable for validating the correctness of the deformation characterization of ISOD DIC. Additionally, metal surfaces exhibit high adhesion to speckled patterns created using the spray paint. This speckled pattern remains stable in the sample without swelling or debonding, ensuring the pixel tracking to find the material deformation. Before yielding, metals exhibit smooth and continuous localized deformation, which leads to low noise during strain validation. As a result, the ISOD-DIC algorithm can be assessed with precision and noise sensitivity. Details of the first experimental case can be found in Section 3.2. 

The second experimental case is a single-edge notched tension (SENT) test for additively manufactured bio-mimetic materials with complex microstructural interfaces. Nacre-inspired brick-and-mortar samples with the size of 7-inch $\times$ 2-inch $\times$ 0.19 inch were printed using a Stratasys J35 polyjet printer. The samples were printed with Verowhite (stiff) and ElasticoBlack (compliant) materials, leading to strain concentration, crack deflection, and a progressive energy dissipation mechanism. This experimental case is designed for the biomimetic samples' highly complex and progressive damage initiation and propagation behaviors, appropriate for demonstrating the effectiveness of ISOD DIC for enhancing the damage characterization data richness. The second experimental case is outlined with details in Section 3.3.
    
By comparing the ISOD DIC results with GOM Correlate results for the two experimental cases, the accuracy and robustness of ISOD DIC will be systematically evaluated and validated.    
    
\subsection{Validation Experimental Case}
\label{sec3.2}

\begin{figure}[htbp]
    \centering
    \includegraphics[width=0.5\textwidth]{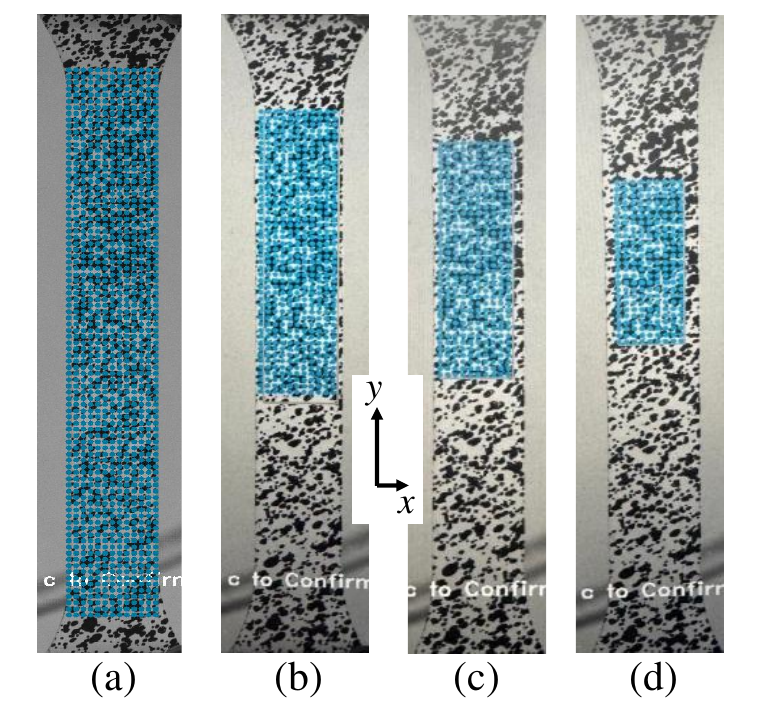}
     \caption{ Aluminum sample with different ROIs: (a) global ROI, (b) ROI 1, (c) ROI 2, (d) ROI 3.  }   
     \label{fig_ROIs}
\end{figure}

Aluminum dog-bone specimens were uniaxially loaded in an MTS load frame with ISOD DIC. A speckled specimen with various ROIs is shown in Figure \ref{fig_ROIs}. Initially, the ROI in Figure \ref{fig_ROIs}(a) was analyzed, and then the ROI was found to be too close to the lateral edges of the specimen. Therefore, a study of the effect of ROI sizes is also conducted in the validation experimental case, as illustrated in Figure \ref{fig_ROIs}. 

\begin{figure}[htbp]
    \centering
    \includegraphics[width=1\textwidth]{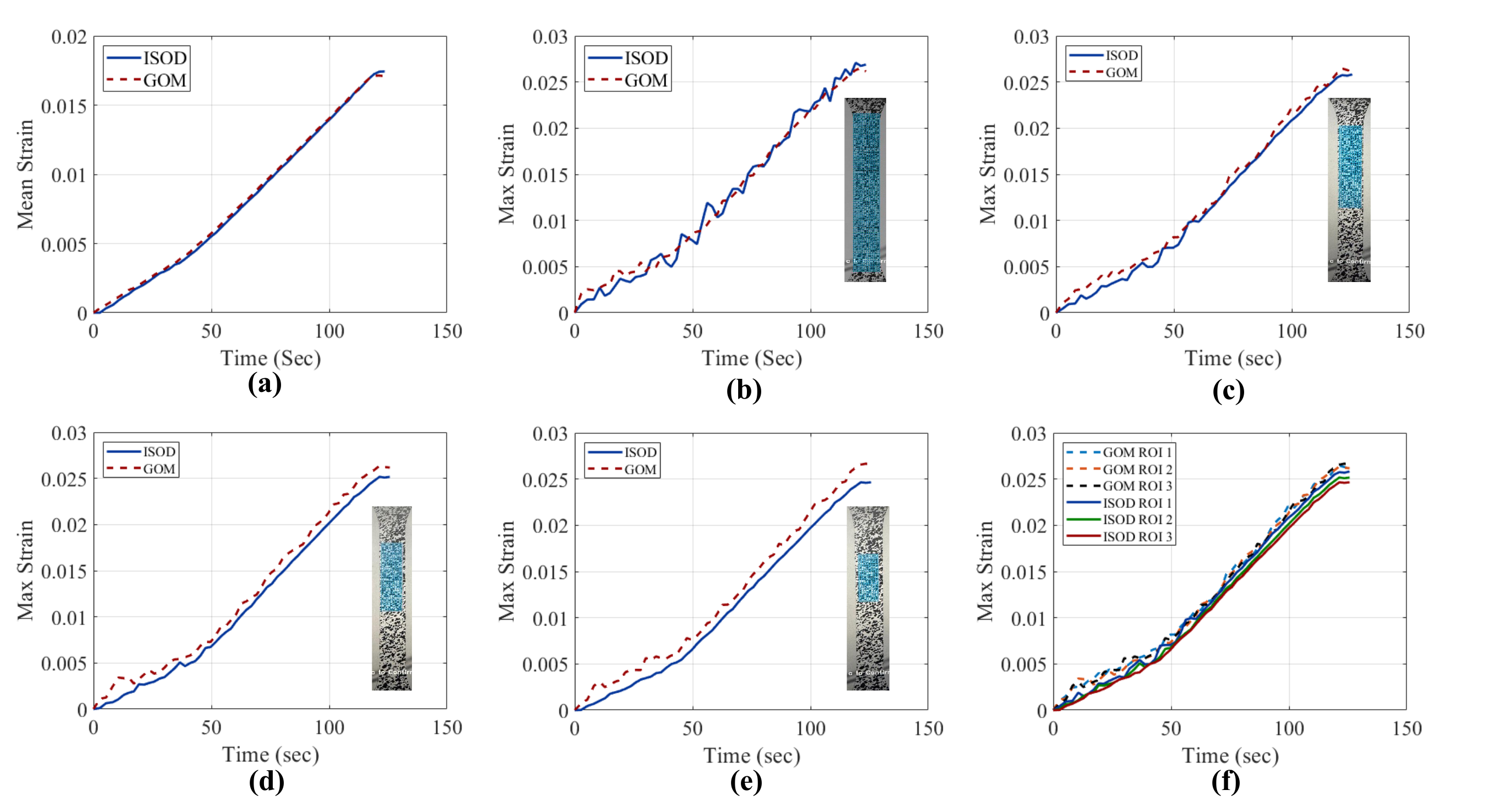}
     \caption{
The history of $\epsilon_y$ (longitudinal strain) measured:
(a) maximum values over the global ROI,
(b) averaged values over the global ROI,
(c) maximum values over ROI~1,
(d) maximum values over ROI~2,
(e) maximum values over ROI~3, and
(f) maximum values comparing the three ROIs with GOM strain data.
}
\label{fig_MAX_Strain}
\end{figure}

The same ROIs were also defined in GOM Correlate to have an accurate and consistent comparison between the ISOD and GOM Correlate DIC measurements. The respective mean strain and maximum strain histories over the global ROI (see Figure \ref{fig_ROIs}(a)) are presented in Figures \ref{fig_MAX_Strain}(a) and (b). In this section, due to the nature of uniaxial loading, only the strain component along the longitudinal direction ($y$ axis) is discussed. The strain component is referred to as the longitudinal strain and  $\epsilon_y$. 

As shown in Figure \ref{fig_MAX_Strain}(a), the longitudinal \textit{MEAN Strain}, which is the average strain of all the elements across the global ROI, evolves linearly throughout the test. The mean strain maintains almost a constant slope from the beginning of the test till the end, indicating a stable and uniform deformation across the sample. The mean strain history curve obtained by ISOD DIC is on top of the curve obtained by GOM Correlate, validating the accuracy of ISOD DIC in terms of global deformation capturing. The longitudinal  \textit{MAX Strain} history over the global ROI is shown in Figure \ref{fig_MAX_Strain} (b). As seen, a good agreement between the ISOD and Gom Correlate DIC results is still seen. However, the ISOD DIC curve exhibits higher oscillations over time. The oscillations in the \textit{MAX Strain}.  curve are mainly influenced by defining the ROI extending to the extreme ends of the specimen. This is due to the optical flow becoming highly sensitive in this region, which leads to unstable displacement and strain approximation with amplified oscillations in \textit{MAX Strain}. 

To mitigate the oscillations in the characterization of \textit{MAX Strain}, three ROIs were defined away from the immediate borders of the specimen. These ROIs are denoted as ROI 1, ROI 2, and ROI 3 in Figures \ref{fig_ROIs}(b) to (d). By excluding the ends of the specimen, where the speckled region is limited, and tracking becomes unstable, the noise in the strain measurements is reduced. The corresponding \textit{MAX Strain} histories of these 3 ROIs are shown in Figure \ref{fig_MAX_Strain} (c)-(e). ISOD-DIC-based \textit{MAX Strain} values are compared with those obtained with GOM Correlate. In addition, Figure \ref{fig_MAX_Strain} (f) shows the direct comparison between the ISOD and GOM Correlate \textit{MAX Strain} histories across all the ROIs. From Figures \ref{fig_MAX_Strain}(c) to (f), it is clear that re-sizing the ROIs away from the edges of the specimen results in more stable and consistent strain histories. \textit{MAX Strain} increases in a linear manner with fewer oscillations. This finding implies the importance of carefully defining the ROI and placing the ROI within the speckled region for smoother strain estimations. 

For all the ROIs, the agreement between ISOD DIC and GOM Correlate remains overall good, as shown in Figure \ref{fig_MAX_Strain} (f) with an approximate error of 0.89\%. Small discrepancies are noticed in Figure \ref{fig_MAX_Strain}(e), with the GOM Correlate curve consistently higher than the ISOD DIC curve. This discrepancy may have been induced by the minor variation of the ROI definitions in ISOD DIC and GOM Correlate. A mismatch of a few columns or rows of pixels may result in a noticeable discrepancy, due to the size of the ROI being relatively small. 
 
\begin{figure}[htbp]
    \centering
    \includegraphics[width=0.9\textwidth]{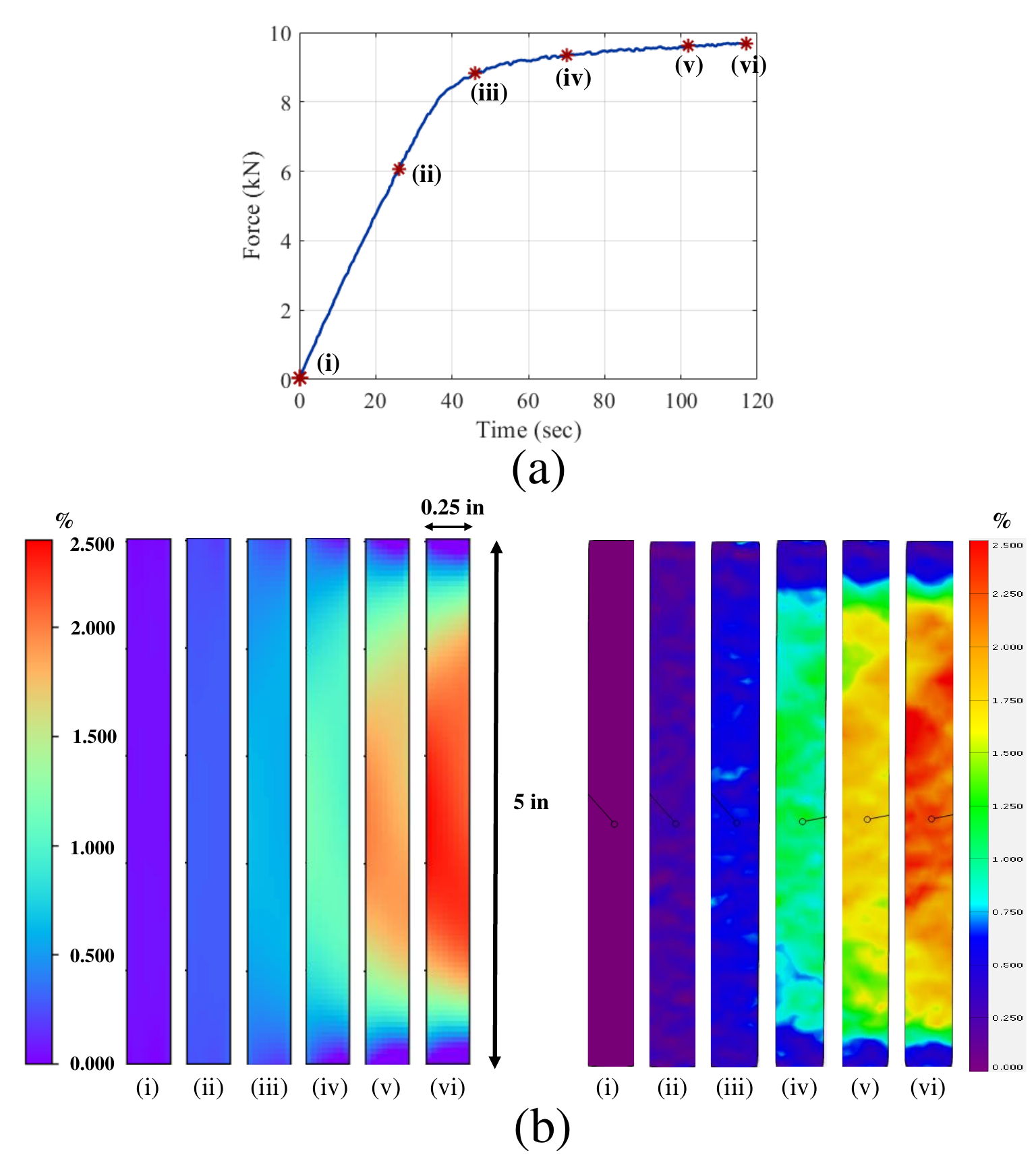}
     \caption{(a) Force-time history with marking corresponding to (b) strain field evaluation of the strain fields at six different stages comparing the ISOD-DIC-generated strain field and the GOM-Correlate-generated strain field.}  
    \label{fig_Strain_Field} 
\end{figure}

To further validate ISOD DIC, strain fields obtained with ISOD DIC and GOM Correlate are compared in Figure \ref{fig_Strain_Field}. The ISOD DIC and GOM Correlate generated longitudinal strain $\epsilon_{y}$ fields across the global ROI at six different stages of the force-time curve, as marked in  Figure \ref{fig_Strain_Field}(a). It is found that the evolution of the strain fields is overall identical, as shown in Figure \ref{fig_Strain_Field}(b), with GOM Correlate results showing more noise. The noisy strain field may be smoothened by built-in spatial filters in GOM Correlate to achieve a better agreement. However, this is not the main goal of this paper and is therefore left out. The good agreement shown in Figure \ref{fig_Strain_Field}(b) further confirms the accuracy of ISOD DIC.

\subsection{Application Experimental Case}
\label{sec3.3}
\begin{figure}[htbp]
    \centering
    \includegraphics[width=1\textwidth]{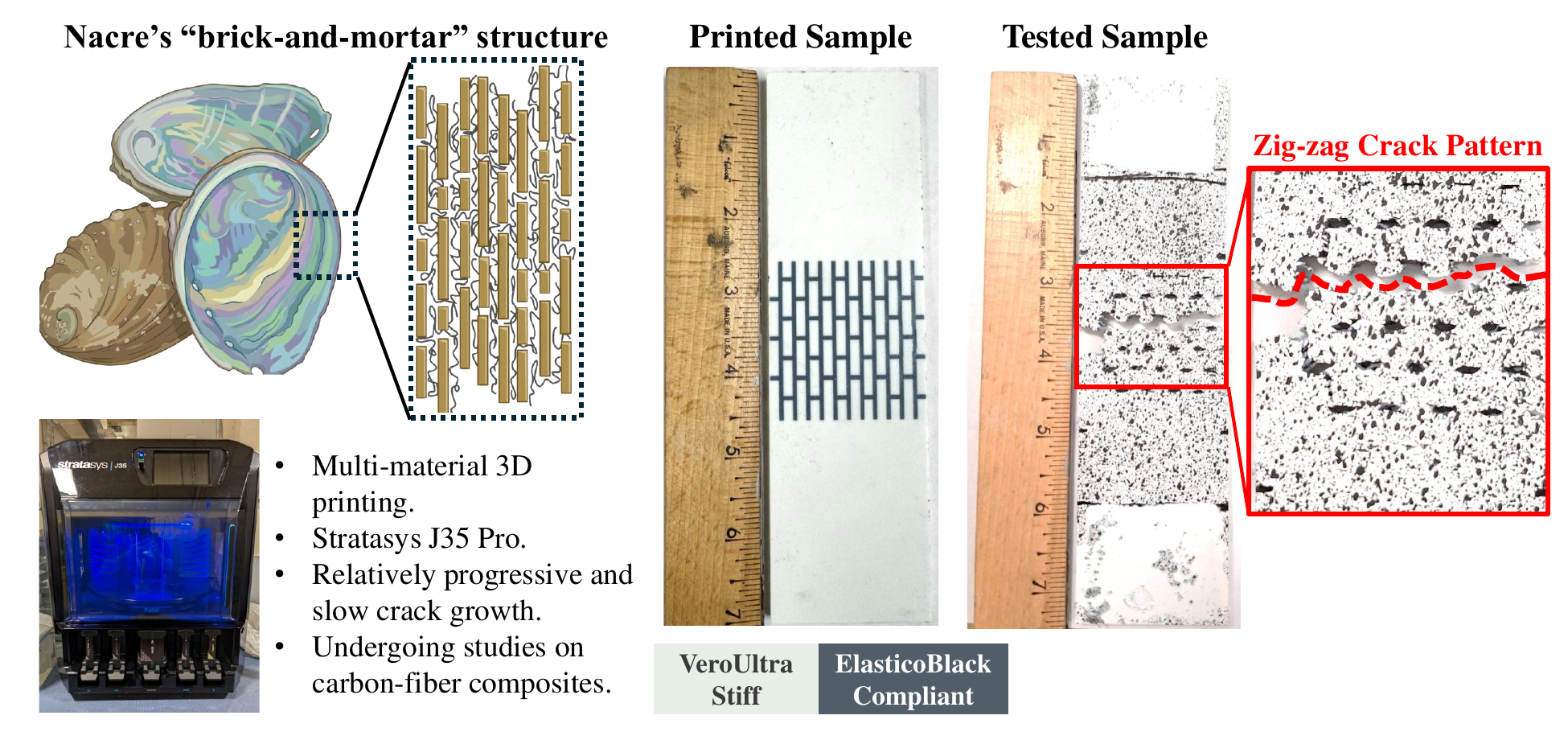}
     \caption{The printed brick-and-mortar sample mimicking nacre's natural microstructure. Stratasys J35 Pro was used for printing the sample, using VeroUltra (stiff) and ElasticoBlack (compliant).}   
     \label{fig_nacre}
\end{figure}

The application experimental case utilizes nacre-inspired additively manufactured specimens. The specimen pattern is also sometimes referred to as brick-and-mortar. The brick-and-mortar structure inspired by nacre's natural microstructure was manufactured using a Stratasys J35 Pro printer with VeroUltra as the stiff material phase and ElisticoBlack as the compliant material phase. A specimen is shown in Figure \ref{fig_nacre}. The specimens were printed with a size of 7 in $\times$ 2 in $\times$ 0.19 in, where the brick-and-mortar pattern extends till 2 inch at the center. The multi-material design of the brick and mortar structure offers superior strength and toughness, as reported in \cite{shao2012nacre1, sun2012nacre2}. As reported, the load-bearing capacity of the brick-and-mortar design offers multiple mechanical advantages. The mortar layer, which acts as a compliant phase, helps in crack deflection and arrests crack propagation rather than passing through the stiff phase. This mechanism facilitates effective crack path length and fracture surface area, resulting in enhanced energy dissipation and delayed failure.

\begin{figure}[htbp]
    \centering
    \includegraphics[width=.9\textwidth]{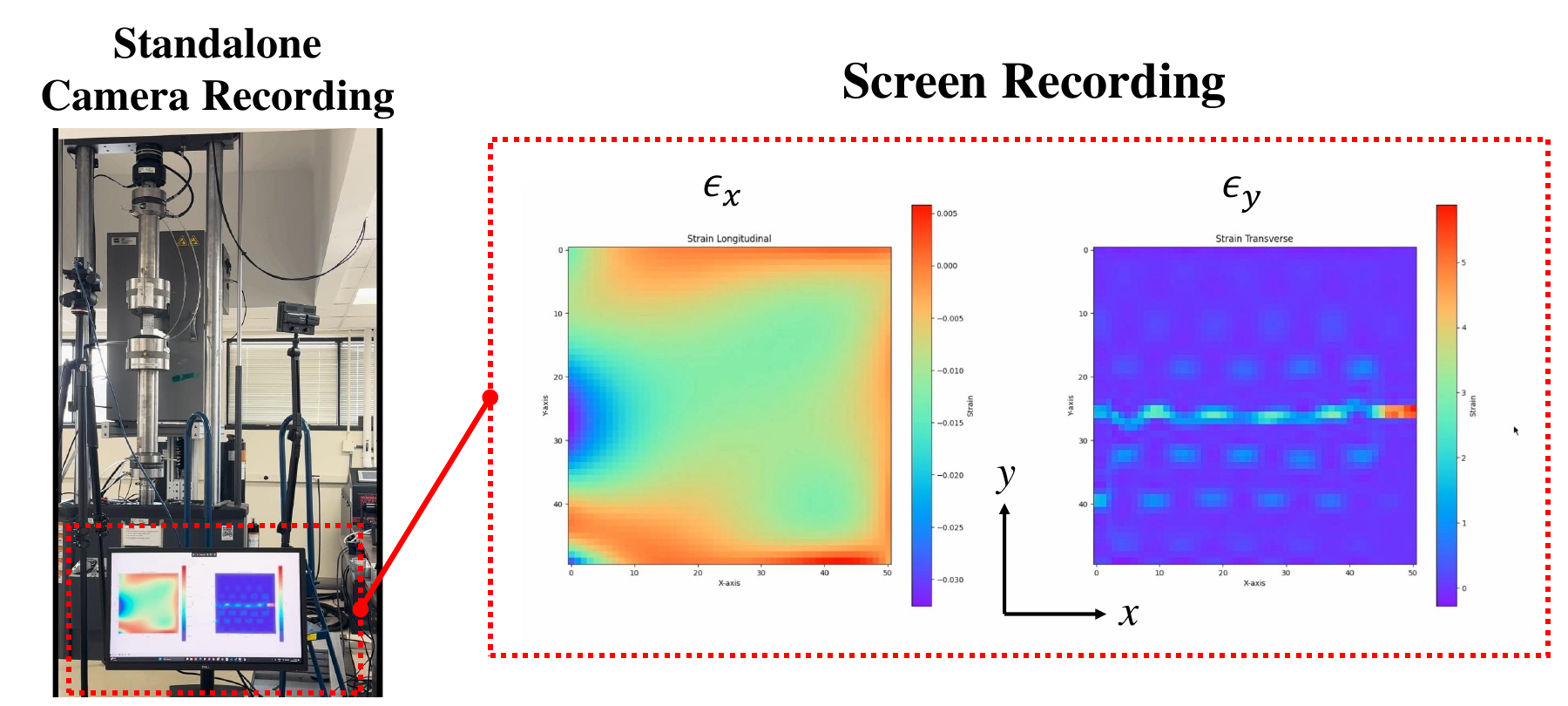}
     \caption{The real-time standalone camera recording and computer screen recording for the testing of the brick-and-mortar samples with ISOD DIC. The camera and screen recordings showcase the capability of ISOD DIC to provide real-time strain field analysis and visualization.}   
     \label{fig_nacre_InSitu}
\end{figure}

ISOD DIC was employed to perform the full-field strain analysis of the brick-and-mortar specimens during the uniaxial tensile loading. The ISOD DIC process flow has been outlined in Figure \ref{fig_ISOD_workflow}. Figure \ref{fig_nacre_InSitu} shows the recordings of the standalone camera and the screen recording of ISOD DIC in-situ characterization. From Figure \ref{fig_nacre_InSitu}, the real-time in-situ DIC measurement provides $\epsilon_{x}$, $\epsilon_{y}$ during the deformation. 
In addition, the strain fields clearly reveal the contrast in strain values between the stiff brick and the compliant mortar phase, demonstrating the capability of the system to clearly capture strain localization across multiple material phases. 

\begin{figure}[htbp]
    \centering
    \includegraphics[width=.72\textwidth]{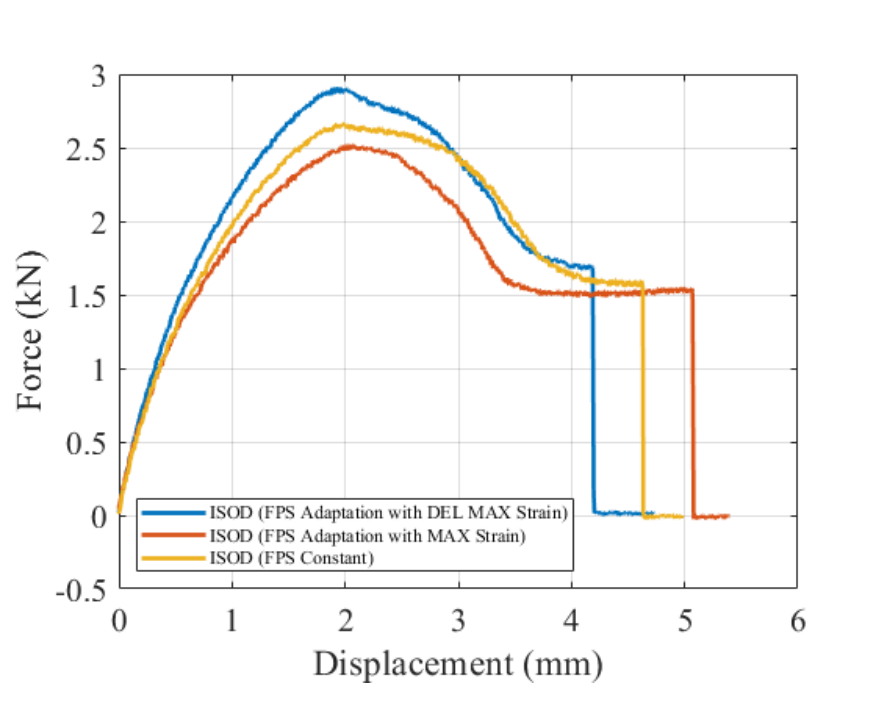}
     \caption{The force-displacement curves of the three samples, demonstrating repeatable mechanical and damage behaviors. The three samples were characterized using ISOD DIC with three camera frame rate updating strategies.}   
     \label{fig_nacre_forceDispCurves}
\end{figure}

The force-displacement curves of the three tested brick-and-mortar specimens are shown in Figure \ref{fig_nacre_forceDispCurves}. The three curves correspond to three samples, loaded with the same mechanical testing settings but with different ISOD DIC strategies to update the camera frame rate. The three strategies include: 1) updating the camera frame rate based on the variation of the maximum strain at the start and end of a imaging batch $\epsilon_{y_{\text{max}}}(t+\Delta t)$-$\epsilon_{y_{\text{max}}}(t)$ (referred to as  \textit{DEL MAX Strain} in Figure \ref{fig_nacre_forceDispCurves}), 2) updating the camera frame rate based on the maximum strain $\epsilon_{y_{\text{max}}}(t)$ (referred to as \textit{MAX Strain} in Figure \ref{fig_nacre_forceDispCurves}), and 3) constant camera frame rate. 
 
Discrepancies in the force-displacement curves are observed in Figure \ref{fig_nacre_forceDispCurves}. The discrepancies are attributed to unavoidable manufacturing defects due to the 3D printing process. Despite the differences, the three curves exhibit consistent mechanical properties, reaching the maximum load in the range of 2.5-2.8 kN and the corresponding displacement in the range of 4.2 -5.1 mm. 

\begin{figure}[htbp]
    \centering
    \includegraphics[width=1\textwidth]{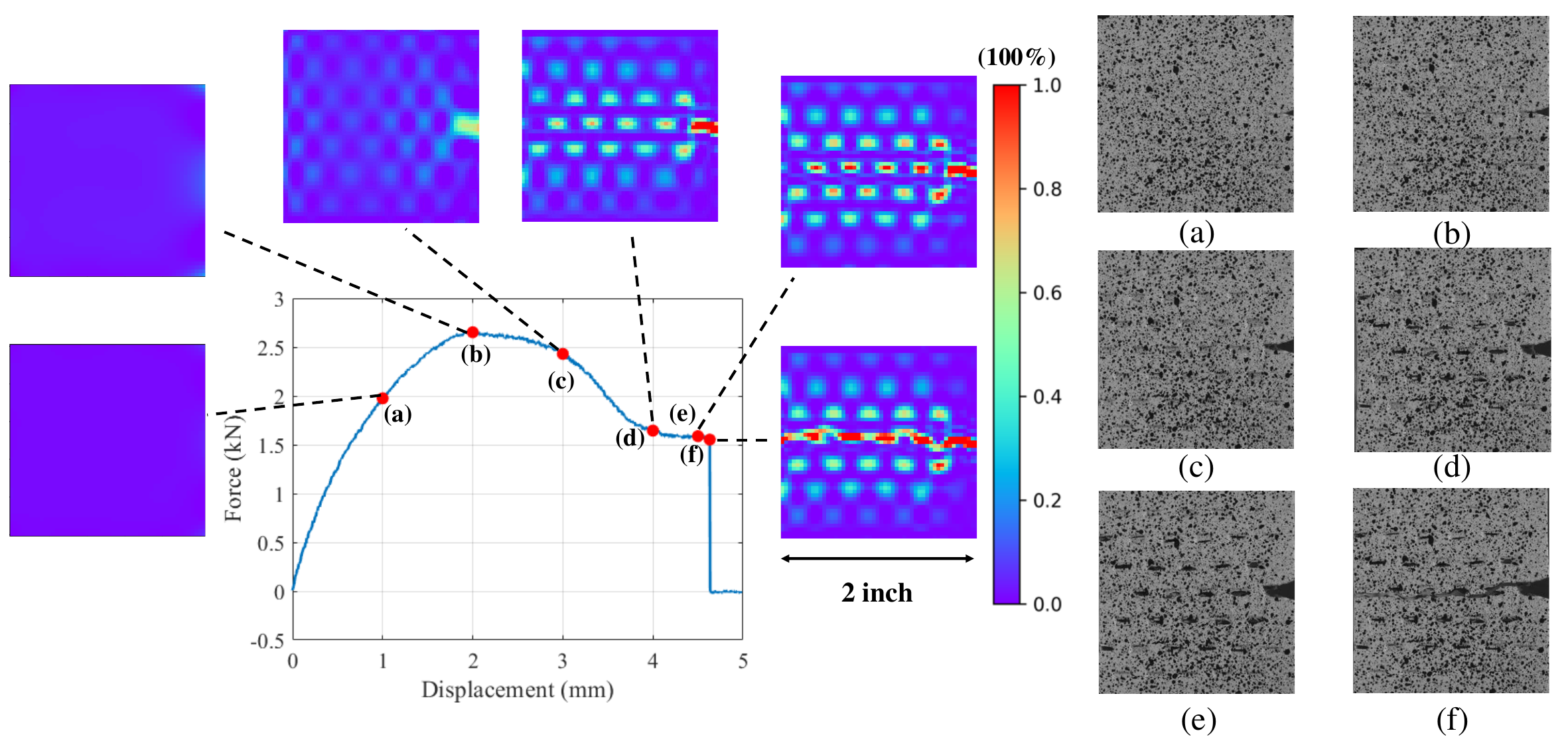}
    \caption{The history of the strain field at 6 different stages across the force vs displacement curve, corresponding raw images shown on the right. }
    \label{fig_nacre_strainHisotry}
\end{figure}

Figure \ref{fig_nacre_strainHisotry} illustrates the damage evaluation of a tested brick-and-mortar specimen. The force-displacement curve on the left is surrounded by the strain field plots representing the damage evolution at 6 different stages until failure, with the corresponding raw speckled images on the right. At Stage (a), before reaching the sample's yield point, the characterized strain field is uniform, indicating elastic deformation. As the force increases from Stage (b) through Stage (d), strain concentration begins to appear at compliant mortar regions. At intermediate stages from Stage (d) to (e), significant strain localization increases in multiple compliant mortar phases, forming a major damage path along the width direction of the sample. At the final Stage (f), the crack path develops from the right end and travels to the left end of the brick-and-mortar specimen, resulting in the catastrophic force drop.  

The images in Figure \ref{fig_nacre_strainHisotry} on the right correspond to raw speckled images at Stages (a) to (f). These images confirm that the damage initiates at various compliant mortar phases instead of starting from a certain location and forming a continuous crack pathway; the crack is arrested near the brick for some time duration. This controlled crack propagation demonstrates the efficiency of the brick-and-mortar sample in delaying catastrophic failure by deflecting cracks and stress redistribution and increasing the overall toughness.

\begin{figure}[htbp]
    \centering
    \includegraphics[width=1\textwidth]{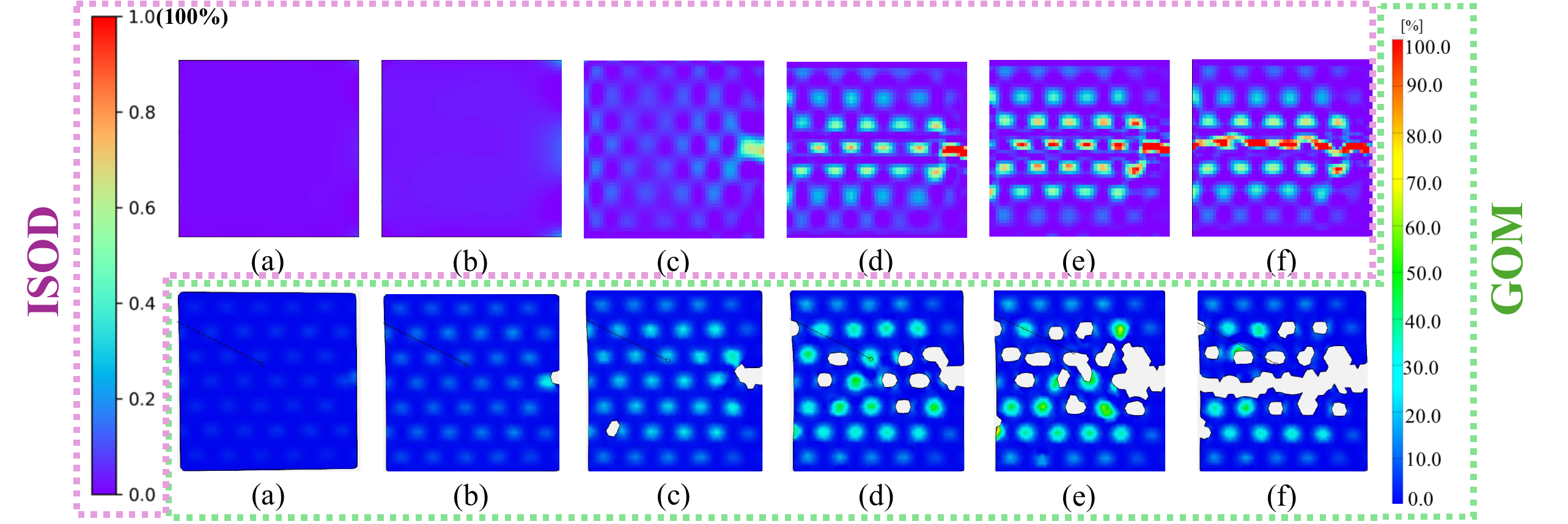}
    \caption{ISOD-DIC-generated strain fields (the upper row) and GOM-Correlate-generated strain fields (the lower row) at the 6 stages described in Figure \ref{fig_nacre_strainHisotry}.   }
    \label{fig_nacre_strainHisotry_vs_GOM}
\end{figure}

The strain fields characterized by ISOD DIC of the 6 stages in Figure \ref{fig_nacre_strainHisotry} are compared with those obtained with GOM Correlate in Figure \ref{fig_nacre_strainHisotry_vs_GOM}. At the early stages from Stage (a) to (c),  strain fields generated by ISOD DIC and GOM Correlate agree well, while the deformation is continuous, before the opening of the cracks. As the test continues, a clear difference can be seen in both the strain fields as the cracks begin to open in the compliant mortar phases, from Stage (d) to (f). In GOM-Correlate-generated strain fields, losses of correlation can be observed, represented by the blank areas in the strain fields. These regions with lost correlation signify excessive local deformation due to crack opening, resulting in missing strain data. Note that correlation tolerance parameters may be adjusted to avoid such losses of correlation in GOM Correlate DIC, but such parameter adjustment may lead to incorrect strain values. In comparison, ISOD-DIC-generated strain fields maintain the correlation and continue to provide insightful data during the crack initiation. Due to the preservation of correlation, the damage initiation and growth history remain clearly visible, demonstrating the robustness of the ISOD DIC in tracking excessive deformation and discontinuities. 

\section{Discussions}
\subsection{Controlling the Camera Frame Rate Based on the Variation of DEL MAX Strain or MAX Strain}

The core idea of ISOD DIC is to integrate a feedback control loop into the DIC characterization process by allowing the extent of deformation to control the image capturing frame rate of the camera. As described in Section 3.3 and Figure \ref{fig_nacre_forceDispCurves}, the camera was controlled according two types of strain measurement: 1) the variation of the max strain at the start and end of a imaging batch $\epsilon_{y_{\text{max}}}(t+\Delta t)$-$\epsilon_{y_{\text{max}}}(t)$ (referred to as \textit{DEL MAX Strain}, and 2) the maximum Strain $\epsilon_{y_{\text{max}}}(t)$ (referred to as \textit{MAX Strain}. The frame rate controlling strategy is essentially that the higher \textit{DEL MAX Strain} or \textit{MAX Strain}  is, the higher the frame rate should be. 

\begin{figure}[htbp]
    \centering
    \includegraphics[width=1\textwidth]{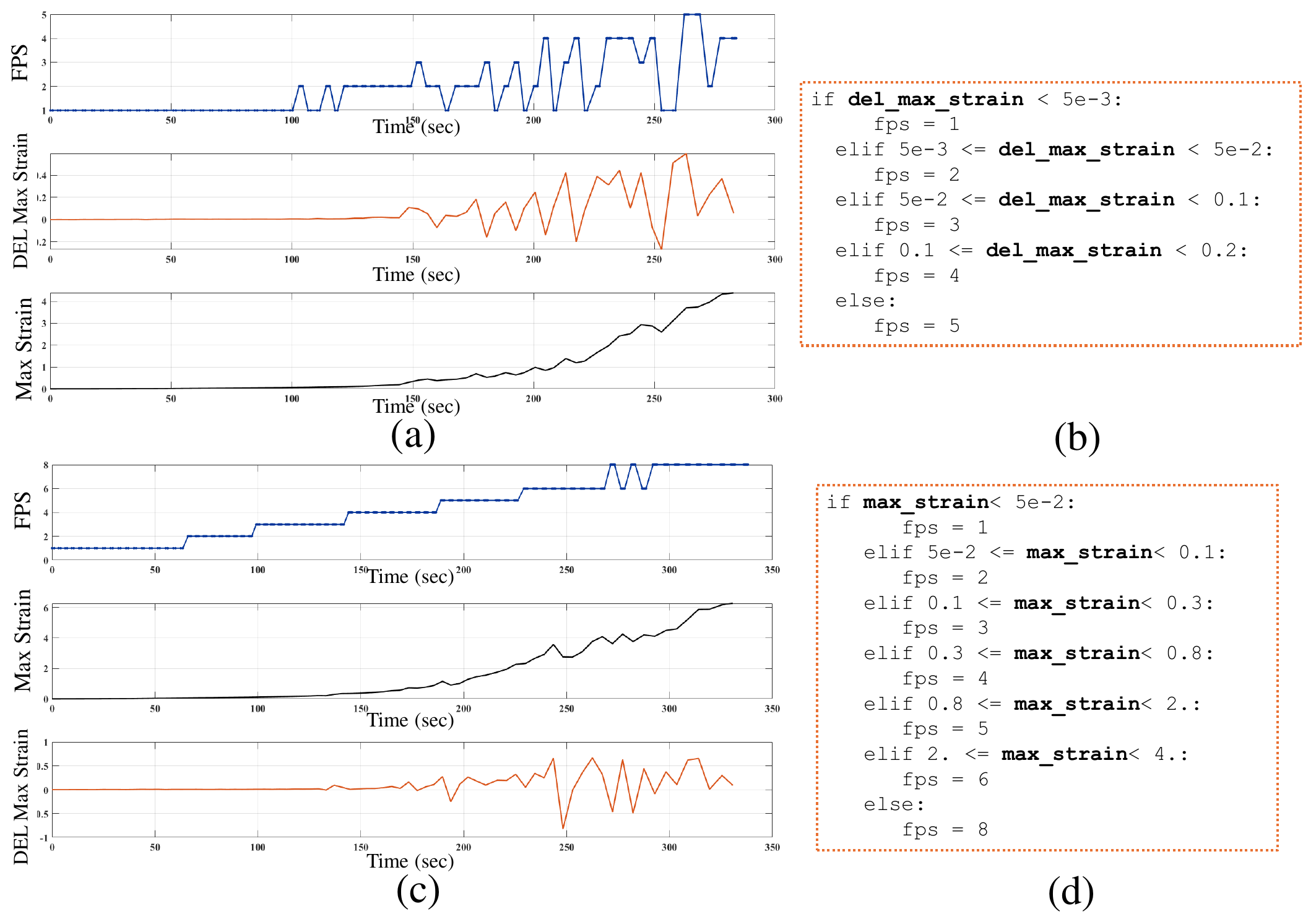}
 \caption{Frame rate adjustment (a) based on \textit{DEL MAX Strain}, 
     (c) based on \textit{MAX Strain}, 
     Threshold limits to adjust the FPS (b) based on \textit{DEL MAX Strain}, and 
     (d)based on \textit{MAX Strain}  } 
     \label{fig_FPS_adaptaion}
\end{figure}

Figure \ref{fig_FPS_adaptaion}(a) illustrates the frame rate adaptation based on \textit{DEL MAX Strain}, with the threshold limits represented in Figure \ref{fig_FPS_adaptaion}(b). In Figure \ref{fig_FPS_adaptaion}(a), the first row records the variation of the camera frame rate over the mechanical test duration. The second row shows the variation of \textit{DEL MAX Strain}, and the third row shows the variation of \textit{MAX Strain}. The frame rate is strictly governed by the rules outlined in Figure \ref{fig_FPS_adaptaion}(b). As seen, \textit{DEL MAX Strain} variation over time is highly oscillatory, directly causing the oscillations in the camera frame rate FPS curve. This behavior becomes evident only after the crack initiation in the compliant phase of the brick-and-mortar specimen. As the crack starts to propagate, the location of the element with the maximum strain shifts frequently. These shifts create oscillations in the \textit{DEL MAX Strain} readings, especially when there are multiple cracks opening and closing. 

On the other hand, Figure \ref{fig_FPS_adaptaion}(c) illustrates the frame rate adaptation based on the \textit{MAX Strain} with the threshold limits shown in Figure \ref{fig_FPS_adaptaion}(d). In Figure \ref{fig_FPS_adaptaion}(c), the first row documents the variation history of the camera frame rate, and the second row shows the history of \textit{MAX Strain}. The third row displays the variation of \textit{DEL MAX Strain}. As the mechanical test goes, \textit{MAX Strain} increases progressively and continuously, and the camera frame rate increases gradually and monotonically. This suggests that using \textit{MAX Strain} as the key parameter to control the camera frame rate appears to be a superior strategy due to its stability and low noise, especially for the experimental cases involving multiple cracks initiating and propagating.  

\begin{figure}[htbp]
    \centering
    \includegraphics[width=1\textwidth]{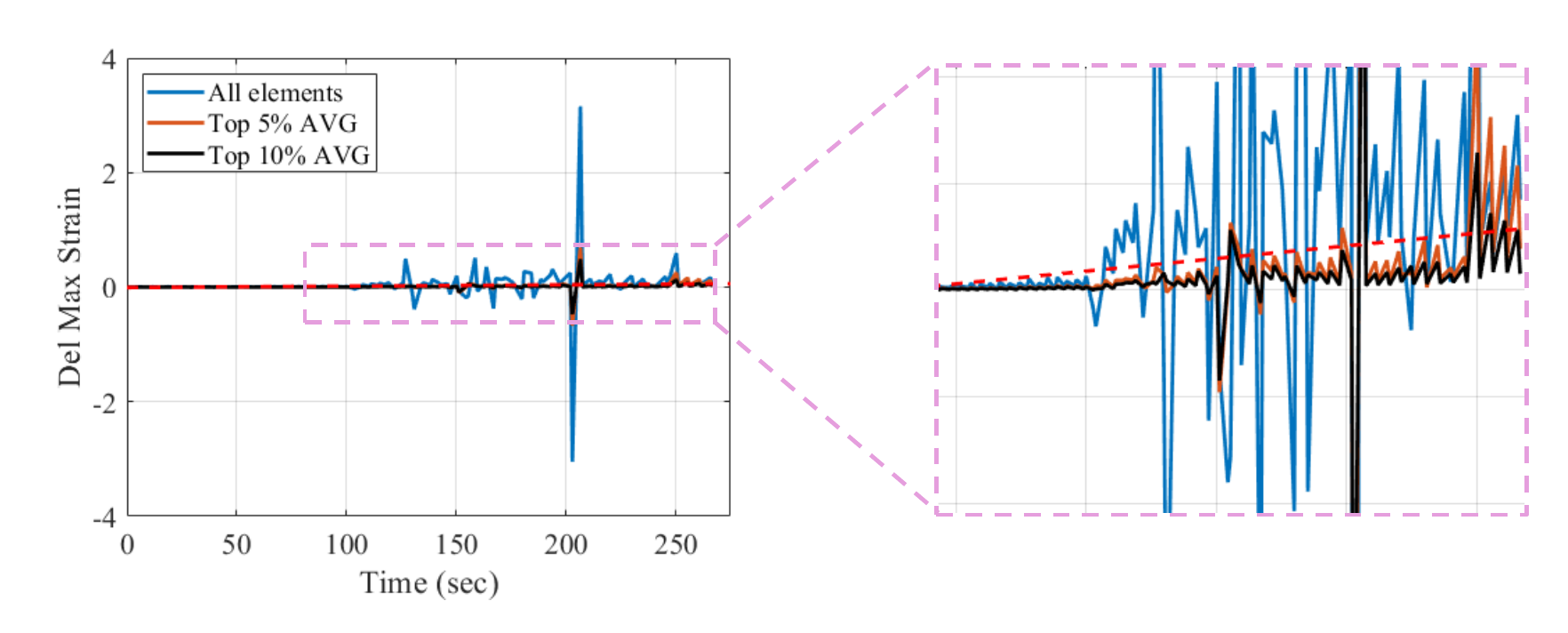}
\caption{The history of \textit{DEL MAX Strain} over the global ROI with various element groups. The top 5\% and 10\% groups of elements are with the highest 5\% and 10\% of the strain values. }
\label{fig_nacer_delmaxstrain}
\end{figure}

A closer look into the oscillatory behavior of the \textit{DEL MAX Strain} curve is shown in Figure \ref{fig_nacer_delmaxstrain}. In this figure, the \textit{DEL MAX Strain} value of all the elements is compared with the sequential variation of the averages taken from groups of elements with the highest 5\% and 10\% strain values. The reason for taking the average values from the groups is to rule out the situation that a few elements with crack opening have singular strain values that should be used as the key parameter to control the camera frame rate. 

As seen in Figure \ref{fig_nacer_delmaxstrain}, when \textit{DEL MAX Strain} of all the elements is considered, the response exhibits high oscillations, causing fluctuations in the camera frame rate adaptation. The response gets smoother by averaging \textit{DEL MAX Strain} by using 5\% and 10\% of all the elements with the highest strain values. With an average of 10 percent elements, the evolution becomes smoother. However, in the locally enlarged view, some oscillations, although smoothened out, still seem unavoidable.

Despite the oscillations over time, the overall growing trend of the \textit{DEL MAX Strain}  across all the cases is preserved. The trend line, shown in Figure \ref{fig_nacer_delmaxstrain} as the dashed curve, indicates the increasing slope as the crack initiates until the catastrophic failure.

\subsection{Information Richness Enhancement due to ISOD DIC}

Table 1 documents the test duration and the number of images captured with various ISOD DIC camera controlling strategies. The three ISOD DIC strategies have been outlined in Section 3.3 and Figure \ref{fig_nacre_forceDispCurves}. An additional test was performed with conventional DIC, meaning that the camera was used during the test with a constant frame rate, without real-time DIC computation or strain field visualization. After the test was done, the captured images were exported to a standalone computer, and GOM Correlate was used for DIC analysis. It should be noted that the lowest camera frame rate of the ISOD DIC characterizations was 1 FPS, while the constant frame rate of the conventional DIC was 1 FPS. 

As shown in Table 1, the test durations of the four tests are similar, but the number of images captured by ISOD DIC and conventional DIC is vastly different. Due to the dynamic camera frame rate adjustment, ISOD DIC controlled by \textit{MAX Strain} captured a high number of 630 images. However, ISOD DIC controlled by \textit{DEL MAX Strain} captured a significantly lower 264 images. The lower image acquisition number of ISOD DIC controlled by \textit{DEL MAX Strain} is due to the oscillatory behaviors of \textit{DEL MAX Strain} and the camera frame rate, as shown in Figure \ref{fig_FPS_adaptaion}. 

In Table 1, although ISOD DIC, with a constant camera frame rate and conventional DIC both used the camera frame rate of 1 FPS, ISOD DIC captured much fewer images, only 50\% of what conventional DIC captured. This discrepancy is due to the fact that ISOD DIC employs real-time in-situ DIC computation and strain visualization, whereas conventional DIC simply performs image acquisition during the test and processes the images with GOM Correlate after the test. 

\begin{table}[htbp]
\centering
\caption{The test duration and number of images captured for different DIC and ISOD configurations.}
\label{tab:test_summary}
\resizebox{0.9\textwidth}{!}{%
\begin{tabular}{lcc}
\hline
\textbf{Test} & \textbf{Test Duration (s)} & \textbf{No. of Images} \\
\hline
ISOD DIC: frame rate based on \textit{MAX Strain}       & 338 & 630 \\
ISOD DIC: frame rate based on \textit{DEL MAX Strain}  & 282 & 264 \\
ISOD DIC: frame rate constant            & 307 & 154 \\
Conventional DIC: frame rate constant                & 309 & 308 \\
\hline
\end{tabular}}
\end{table}

Table 2 provides more insights by separating the test into two phases, including the minimal crack growth phase from Stage (a) to Stage (b) and the significant crack growth phase from Stage (b) to Stage (f), as illustrated in Figure \ref{fig_nacre_strainHisotry}. Table 2 highlights the effectiveness of ISOD DIC. For ISOD DIC controlled by \textit{MAX Strain}, at the minimal crack growth phase ( lasting 123 s), 114 images were acquired with real-time deformation assessment and visualization. At the significant crack growth phase (lasting 215 s), 516 images were acquired with real-time analysis and visualization. As shown in Table 2, although the significant crack growth phase is only 75\% longer than the minimal crack growth phase, the number of images acquired increases by 353\%. This contrast clearly shows the effectiveness of ISOD DIC, which is that more images should be taken when events of interest happen (e.g., yielding, damage, etc.), allowing for enriched information acquisition. ISOD DIC enables detailed tracking of the crack initiation and propagation, signified by localized strain evolution. 

\begin{table}[htbp]
\centering
\caption{Comparison of time duration and number of images captured during minimal and significant crack growth for different testing strategies.}
\label{tab:crack_growth_comparison}
\resizebox{\textwidth}{!}{%
\begin{tabular}{lcccc}
\toprule
\multirow{2}{*}{\textbf{Test}} 
& \multicolumn{2}{c}{\textbf{Minimal crack growth}} 
& \multicolumn{2}{c}{\textbf{Significant crack growth}} \\
\cmidrule(lr){2-3} \cmidrule(lr){4-5}
& \textbf{Time Duration (s)} 
& \textbf{No. of Images} 
& \textbf{Time Duration (s)} 
& \textbf{No. of Images} \\
\midrule
ISOD DIC: frame rate based on \textit{MAX Strain}        & 123 & 114 & 215 & 516 \\
ISOD DIC: frame rate based on \textit{DEL MAX Strain} & 117 &  67 & 165 & 197 \\
ISOD DIC: frame rate constant                                  & 119 &  66 & 188 &  88 \\
\bottomrule
\end{tabular}}
\end{table}

There are still several limitations of ISOD DIC. First, maintaining a constant light setup is essential for optical flow measurement because optical flow assumes constant light intensity throughout the test. Significant fluctuations in the light intensity may lead to inaccurate displacement field estimation and reduce the reliability of the results. Therefore, ensuring the stable light intensity throughout the test is crucial when capturing images. Another limitation arise from camera activation. After each batch of imaging, as shown in Figure \ref{fig_Batch_wise_DIC}, there will be a small dormant time duration (e.g., $x$ s and $y$ s in Figure \ref{fig_Batch_wise_DIC}) for the activation of the machine vision camera. Such dormant time is usually around 1 second and may be minimized by adjusting camera settings, using high-speed connectors, and faster CPUs.  

\section{Conclusions}
\label{Section5}
A DIC paradigm, in-situ on-demand (ISOD) DIC, has been developed and presented in this paper. The core idea of ISOD DIC is to integrate the camera frame rate control into the DIC process flow, allowing for dynamic adjustment of the frame rate according to the deformation of the inspected sample. With higher local deformation and deformation rate, the integrated camera captures images with an elevated frame rate, enabling enhanced data richness for events of interest, such as material yielding, creeping, and crack development. 

2D ISOD DIC has been implemented and validated in this paper. First, the accuracy of ISOD DIC was verified by performing uniaxial tensile tests on aluminum dog-bone samples. The strain fields computed by ISOD DIC and a commercial package, GOM Correlate, were compared and found to agree well, with the error of 0.89\%. Then, ISOD DIC was applied to provide real-time characterization for the deformation and cracking of additively manufactured biomimetic specimens due to single-edge notched tension loading. ISOD DIC was found to be able to noticeably enhance the information richness by dynamically increasing the frame rate and by capturing  516 images, which is approximately 178\% more than the images captured by conventional DIC, i.e., 188 images, especially for the mechanical loading phase with significant damage development. This means that the new ISOD DIC paradigm is able to significantly increase the data availability for the events of interest, such as yielding, damage, creeping, etc., without consuming excessive storage space and analysis time. 

Compared to conventional DIC, the computation core of ISOD DIC is a Lucas–Kanade (LK)-based optical flow model. The two prominent advantages of ISOD DIC are that it enables real-time DIC analysis and visualization, and that the camera can be dynamically controlled to acquire images faster when the inspected specimen deformation becomes excessive and fast-growing. ISOD DIC, similar to other DIC models based on optical flow, is sensitive to environmental lighting changes and needs to be performed with constant lighting conditions. Future works include making ISOD DIC more robust against environmental lighting variations and leveraging the closed-loop controlling capability of ISOD DIC to govern the operation and manipulation of other sensors, such as thermography cameras. 

\section*{Acknowledgments}
The authors gratefully acknowledge the financial support from the U.S. Air Force Office of Scientific Research (AFOSR) Defense University Research Instrumentation Program (DURIP). 

\bibliographystyle{elsarticle-num}
\bibliography{sample}







\end{document}